**General expression for the energy**
**and the equation of state for solids**

Bystrenko O., Ilkiv B., Petrovska S., Bystrenko T., Foia O., Khyzhun O.

Frantsevich Institute for Problems of Materials Science
National Academy of Science of Ukraine

Abstract: On the basis of the extended classical elasticity theory, we propose universal semi-empirical analytical expressions for the energy and the equation of state (EOS) for polycrystalline solids. The validation of the relations is made by means of first principle DFT simulations with the use of pseudopotential approach and GGA approximation for the exchange-correlation energy. The calculations performed for a large number of inorganic crystalline compounds with metal, covalent and ionic bonding, including diamond C (Fd-3m), α-Mg ($P6_3$/mmc), ZnS (F-43m), α-$B_{12}$ (R-3m), $MgC_2B_{12}$ (Imma), $NaAgO_2$ (C2/m), etc., within the pressure range up to ~300 GPa demonstrate an excellent agreement with the predictions of the analytical theory comparable in accuracy with Birch-Murnaghan approach.

## Introduction

Equations of state (EOS) for solids attract the attention of researchers in view of their fundamental meaning and applications, for instance, in geophysical science, where it is important to understand the behavior of solids under high pressure. At present, a large number of empirical and semi-empirical EOS is known (see, for instance, Refs.[1-8]). In particular, approaches proposed by Birch – Murnaghan [1, 2], Grüneisen [3], Holzapfel [4], Vinet [5] are frequently used.

Purely empirical approaches are mostly accurate for some classes of compounds, while those employing thermodynamical considerations represent the free energy and pressure in the form of power series of strain parameters, requiring, consequently, higher powers to be included to achieve better accuracy. Typically, (semi-) empirical EOS establish the relation P=P(V) for zero temperature, and the general isotherms P=P(V,T) are obtained by introducing the temperature dependence for relevant parameters, or by adding some entropy related terms to the expressions for the free energy.

In this work, we propose semi-empirical expressions for energy and related equation of state for solids, which have a very simple functional form reflecting their relation to the classical elasticity theory for small deformations. At the same time, these expressions can be considered universal, since they accurately reproduce the behavior of the elastic energy and pressure within wide range of strains for various types of compounds, with different type of crystal systems and bonding (i.e., for metals, ionic and covalent compounds). The validation of the analytical relations is made by means of numerical simulations on the basis of first principle density functional theory (DFT) approach, which has proven to be an accurate tool for the description of structural and mechanical properties of inorganic compounds.

## Theoretical background

Let us begin with recalling some basic relations of classical elasticity theory. The density of elastic energy of an isotropic solid in the case of small deformations is given by the relation [9]

$$W=\frac{B}{2}\cdot I_B{}^2+G\cdot\sum_{ij}^{3}\left(\varepsilon_{ij}-\frac{1}{3}\delta_{ij}I_B\right)^2. \qquad (1)$$

Here the quantities

$$I_B=\sum_{i}^{3}\varepsilon_{ii}\ , \qquad (2)$$

and

$$I_G=\sum_{ij}^{3}\left(\varepsilon_{ij}-\frac{1}{3}\delta_{ij}I_B\right)^2 \qquad (3)$$

are two independent scalar invariants associated with the symmetric second-rank strain tensor $\varepsilon_{ij}$. Relations (1-3) define two independent elastic parameters of an isotropic solid. In particular, the coefficient B is the bulk modulus.

In what follows, we consider non-shear volumetric (hydrostatic) compression (or expansion) of isotropic polycrystalline solids only, therefore, $I_G=0$, and Eq. (1) simplifies to

$$W=\frac{B}{2}\cdot I_B^2\ , \qquad (4)$$

where the respective isotropic bulk modulus can be found, for instance, from the Voigt- Reuss-Hill relations [10].

Notice, that under small volumetric compression of an isotropic medium, the non-zero components of the strain tensor are equal, $\varepsilon=\varepsilon_{11}=\varepsilon_{22}=\varepsilon_{33}$ , therefore,

$$V=V_0(1+\varepsilon_{11})(1+\varepsilon_{22})(1+\varepsilon_{33})=V_0(1+\varepsilon)^3 \quad (5)$$

On this basis, we can introduce the following strain parameter,

$$\xi=(1-x)\simeq-3\varepsilon=-I_B, \quad (6)$$

where

$$x=V/V_0. \quad (7)$$

Then, for the elastic energy of a solid of volume $V_0$ for small strains we can rewrite Eq. (4) as

$$U=\frac{1}{2}B_0V_0\xi^2. \quad (8)$$

To point out that the bulk modulus is measured at zero pressure, we use hereafter the notation $B_0$.

We emphasize that the relation (8) is accurate for small deformations of isotropic media. Actually, it may be regarded as another formulation of Hooke’s law. However, from the above mentioned, one can suppose that for large (non-linear, but elastic) strains the expression for the elastic energy should look as

$$U=\frac{1}{2}B_0V_0\xi^2\Phi(\xi) \quad (9)$$

where $\Phi(\xi)\to1$ as $\xi\to0$.

The basic empirical assumption of this work is that the elastic energy of an isotropic polycrystalline solid can be, in general, described by the function

$$U=\frac{1}{2}B_0V_0\xi^2\frac{1}{(1-\alpha\xi)^2}, \quad (10)$$

i. e., we accept that $\Phi(\xi)=1/(1-\alpha\xi)^2$. Notice that it has the form (9).

Then, the relevant EOS for zero temperature can be derived as

$$p(\xi)=-\frac{\partial U}{\partial V}=\frac{B_0\xi}{(1-\alpha\xi)^3} \quad (11)$$

Here the quantity α can be regarded just as a fitting parameter. However, its value can be related as well to the first derivative of bulk modulus with respect to pressure from thermodynamic considerations, similar to how it is done when deriving the Birch-Murnaghan (BM) EOS. Namely, given the relation

$$B=-V\frac{\partial P}{\partial V} \quad , \tag{12}$$

we can write

$$B'=\frac{\partial B}{\partial P}=-\frac{\partial B}{\partial V}\frac{V}{B}, \tag{13}$$

and then, by using Eqs. (10-11), express the value of α in terms of the derivative B′ in the limit $\xi \rightarrow 0$, i.e. for the pressure p=0. After performing a lot of cumbersome calculations, one can arrive at the following result:

$$\alpha=\left(B_0{}'+1\right)/6. \tag{14}$$

Let us recall that 3-rd order BM approach similarly contains 2 parameters, bulk modulus $B_{0BM}$ and its first derivative $B_0'_{BM}$ (with respect to pressure); the respective expressions for the energy and pressure are

$$U_{BM}=\frac{9V_0 B_{0BM}}{16}\left[\left(x^{-2/3}-1\right)^3 B_0{}'_{BM}+\left(x^{-2/3}-1\right)^2\left(6-4x^{-2/3}\right)\right] \tag{15}$$

$$p_{BM}=\frac{3B_{0BM}}{2}\left(x^{-7/3}-x^{-5/3}\right)\left[1+\frac{3}{4}\left(B_0{}'_{BM}-4\right)\left(x^{-2/3}-1\right)\right], \tag{16}$$

where x is defined by Eq.(7).

## Results of numerical validation

Validation of the relations (10-11) has been done on the basis of first principle density functional theory (DFT) as implemented in Quantum Espresso software [11]. In simulations, the standard solid state pseudo-potential library SSSP, precision version 1.1.2 [12], with the exchange-correlation energy given in Perdew–Burke-Ernzerhof form [13] was employed. The sampling of the Brillouin zone was made on the uniform Monkhorst-Pack grid [14], predominantly of the size 8x8x8, with no shift. The cutoff energies for the wave functions and charge were set according to the systems considered and, respectively, the pseudo-potentials used; the total energy convergence threshold for the electron wave functions was set $10^{-7}$ Ry.

Initial crystallographic information for the compounds considered was taken from the materials database [15]. Then, in all cases, the simulations were preceded by optimization (relaxation) of initial structures. Optimization was performed with the accuracy of $10^{-4}$ Ry/Bohr by means of BFGS method [16] and was aimed at finding the equilibrium ion configurations and lattice parameters associated with the minimum energy at zero pressure.

The model which we employed for the evaluation of the energy and pressure of strained solids can be regarded as an extension of the approach used by Voigt for calculation of isotropic elastic moduli of polycrystalline aggregates [17] for the case of large deformations. This means that we neglect the inter-grain interaction and possible eigenstrains and assume that the strain tensor remains constant over all polycrystalline structure after applying finite volumetric compression. In this case all grains are deformed uniformly and the energy per atom and pressure can be determined from the DFT simulation performed for a single strained supercell, subject to ionic relaxation. The consideration is limited to the maximum strain $\xi_{max}$~0.4, since the use of pseudo-potentials may cause errors at extremely high densities.

Numerical simulations were performed for a large number of inorganic compounds with various crystal systems and space symmetry, and with metal, ionic, and covalent bonding, in particular, Cu (Fm-3m); Au (Fm-3m); α-Mg ($P6_3$/mmc); $Cu_3Au$ (Pm-3m); NaCl (Fm-3m); $CaF_2$ (Fm-3m); $Al_2O_3$ (R-3c); ZnS (F-43m); ZnS (Fm-3m); diamond C (Fd-3m); α-$B_{12}$ (P-3m); γ-$B_{28}$ (Pnnm); $B_4$ (Cmce); $TiB_2$ (P6/mmm); $MgC_2B_{12}$ (Imma); topaz $Al_2SiO_4F_2$ (Pnma); etc.

In order to elucidate the maximum accuracy potential of the proposed model, both bulk moduli $B_0$ and the parameters α needed for the relations (10-11) were evaluated for each compound on the basis of DFT simulations, by minimizing the mean relative error (MRE) for pressure

$$R=\frac{1}{M}\sum_{k=1}^{M}\left|\left[P_k-p\left(\xi_\kappa\right)\right]/P_k\right| \ , \qquad (17)$$

where M is the number of strain points used in DFT calculations (in most cases, it was equal M=15); $P_k$ and $p(\xi_k)$ are, respectively, the pressure obtained in simulation and that given by Eq.(11) at the point $\xi_k$. In addition, a similarly defined MRE for the energy per particle

$$S=\frac{1}{M}\sum_{k=1}^{M}\left|\left[\Delta U_k-U\left(\xi_\kappa\right)\right]/\Delta U_k\right| \qquad (18)$$

was calculated. Here $\Delta U_k$ is the energy obtained in DFT calculations for the point $\xi_k$ with respect to the ground state energy at zero pressure (for ξ=0), while U(ξ) is given by Eq.(10) .

In addition, accuracy of the relations (10)-(11) has been compared with the 3-rd order BM approach. The respective evaluation of bulk moduli $B_{0BM}$ and their derivatives $B_0'_{BM}$ with respect to pressure was performed independently, in the same way, i.e., by minimizing MRE (17) for the pressure given by the BM expression (16).

The results of calculations for the energy and pressure for a typical example, $B_6O$ (R-3m) compound, are given in Fig. 1. More detailed data for a number of selected inorganic compounds with different crystal systems and types of bonding can be found in Table 1.

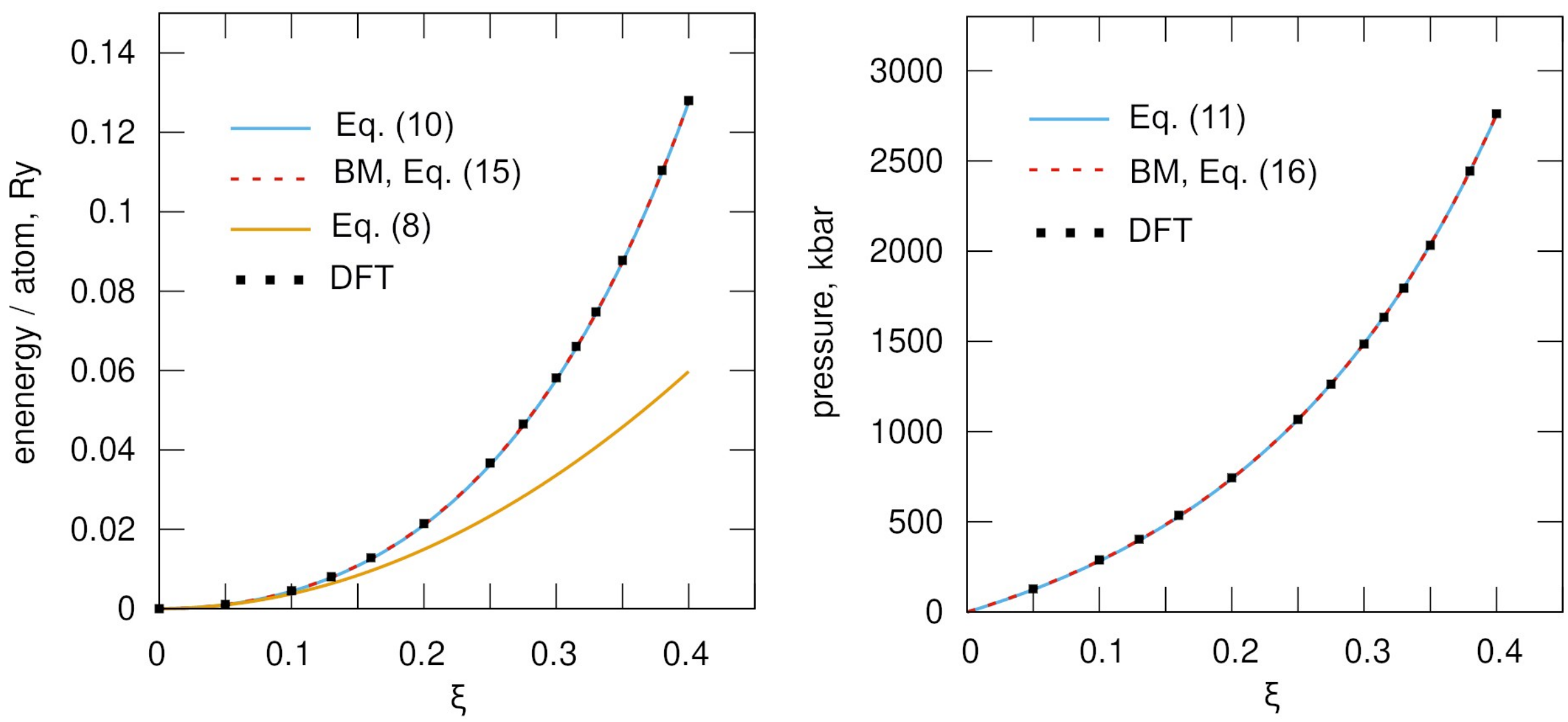

Fig. 1. Comparison of the results of first principle DFT calculations for energy and pressure with the theoretical approaches (10-11), BM (15-16), and (8) for the compound $B_6O$ (R-3m).

As is seen from the figure, both BM relations (15-16) and Eqs. (10-11) perfectly reproduce the numbers obtained in DFT calculations. It is clear that a such a good agreement is the result of the independent fitting of bulk moduli and other parameters needed for these two approaches. This demonstrates, on the one hand, the relative adequacy of the model of solids used for numerical validation, and, on the other hand, the potential accuracy of the theoretical models considered.

Table 1. Bulk moduli and related parameters for a number of inorganic compounds and MRE for energy and pressure calculated for the relations (10-11) and BM (15-16) with respect to first principle DFT results.

| No | Compo-sition | Crystal system, symmetry, space group number | Bulk modulus $B_0$ , $B_{0BM}$ , (kbar) | Parameters $\alpha$, $B_0'_{BM}$ | MRE for the energy, Eqs.(10) vs. BM (15) | MRE for the pressure, Eqs.(11) vs. BM (16) |
|---|---|---|---|---|---|---|
| 1 | Diamond, C | Cubic,<br>Fd-3m [227] | 4321.4<br>4368.5 | 0.800<br>3.628 | $9.0\cdot10^{-4}$<br>$7.7\cdot10^{-4}$ | $1.7\cdot10^{-3}$<br>$4.1\cdot10^{-4}$ |
| 2 | ZnS | Cubic,<br>F-43m [216] | 707.1<br>715.1 | 0.908<br>4.252 | $1.2\cdot10^{-3}$<br>$1.3\cdot10^{-3}$ | $8.9\cdot10^{-3}$<br>$6.6\cdot10^{-3}$ |
| 3 | $CaF_2$ | Cubic,<br>Fm-3m [225] | 770.7<br>803.2 | 0.933<br>4.293 | $5.3\cdot10^{-5}$<br>$4.4\cdot10^{-4}$ | $3.7\cdot10^{-4}$<br>$3.3\cdot10^{-3}$ |
| 4 | $\alpha$-Mg | Hexagonal,<br>$P6_3$/mmc [194] | 358.4<br>362.3 | 0.851<br>3.914 | $9.4\cdot10^{-3}$<br>$9.5\cdot10^{-3}$ | $5.9\cdot10^{-3}$<br>$8.1\cdot10^{-3}$ |
| 5 | $TiB_2$ | Hexagonal,<br>P6/mmm [191] | 2513.6<br>2545.8 | 0.836<br>3.820 | $5.6\cdot10^{-3}$<br>$4.9\cdot10^{-3}$ | $1.3\cdot10^{-3}$<br>$2.0\cdot10^{-3}$ |
| 6 | $\alpha$-B | Trigonal,<br>R-3m [166] | 2123.8<br>2146.1 | 0.795<br>3.599 | $8.2\cdot10^{-4}$<br>$7.9\cdot10^{-4}$ | $1.0\cdot10^{-3}$<br>$1.0\cdot10^{-3}$ |
| 7 | $B_6O$ | Trigonal,<br>R-3m [166] | 2259.3<br>2277.7 | 0.769<br>3.481 | $1.3\cdot10^{-4}$<br>$1.0\cdot10^{-4}$ | $4.8\cdot10^{-3}$<br>$3.2\cdot10^{-3}$ |
| 8 | $Al_2O_3$ | Trigonal,<br>R-3c [167] | 2312.1<br>2342.6 | 0.864<br>3.981 | $2.9\cdot10^{-4}$<br>$2.5\cdot10^{-4}$ | $1.8\cdot10^{-3}$<br>$8.7\cdot10^{-4}$ |
| 9 | $MgC_2B_{12}$ | Orthorhombic,<br>Imma [74] | 2294.6<br>2313.2 | 0.790<br>3.589 | $4.2\cdot10^{-4}$<br>$4.1\cdot10^{-4}$ | $1.1\cdot10^{-3}$<br>$6.3\cdot10^{-4}$ |
| 10 | $Al_2SiO_4F_2$ | Orthorhombic,<br>Pnma [62] | 1412.8<br>1427.4 | 0.842<br>3.860 | $4.1\cdot10^{-4}$<br>$3.5\cdot10^{-4}$ | $1.3\cdot10^{-2}$<br>$1.0\cdot10^{-2}$ |
| 11 | $NbAl_3$ | Tetragonal,<br>I4/mmm [139] | 1266.7<br>1298.0 | 0.855<br>3.878 | $1.5\cdot10^{-3}$<br>$1.4\cdot10^{-3}$ | $9.5\cdot10^{-4}$<br>$1.7\cdot10^{-3}$ |
| 12 | $NaAgO_2$ | Monoclinic,<br>C2/m [12] | 643.9<br>645.6 | 0.991<br>4.902 | $2.5\cdot10^{-4}$<br>$2.5\cdot10^{-4}$ | $7.4\cdot10^{-4}$<br>$1.3\cdot10^{-3}$ |
| 13 | $CaP_3$ | Triclinic,<br>P-1 [2] | 500.57<br>512.57 | 0.847<br>3.833 | $6.6\cdot10^{-4}$<br>$3.7\cdot10^{-4}$ | $4.7\cdot10^{-3}$<br>$2.2\cdot10^{-3}$ |

The numbers for relative errors given in Table indicate that the proposed relations (10-11) are on average comparable in accuracy with 3-rd order BM approach, demonstrating slightly better results for the compounds with $\alpha \gtrsim 0.9$ ($B_0'_{BM} \gtrsim 4.3$). At the same time, the accuracy of BM (15-16) and (10-11) relations correlates for different compounds, which means that the simple relations (10-11) can be regarded just as a good approximation for BM equations.

## Conclusions

The proposed simple semi-empirical relations (10) and (11) perfectly reproduce the results of first principle calculations of the energy and pressure of solids in a wide strain and pressure range ($\xi_{max}$~0.4, $P_{max}$~ 3 Mbar) for a large number of inorganic solids with different crystal systems and type of chemical bonding (covalent, ionic and metallic), with the accuracy comparable with 3-rd order Birch-Murnaghan theory. The Birch-Murnaghan theory better reproduces the properties of compounds with lower values of the bulk modulus derivative, while the approach (10-11) is slightly more accurate for $B_0'_{BM} \gtrsim 4.3$.

## Acknowledgment

The authors thank Prof. A. Zagorodny for the interest in the work and valuable discussions.